\documentclass[%
reprint,
superscriptaddress,
twocolumn,
notitlepage,
amsmath,amssymb,
aps,
pra,
]{revtex4-1}

\usepackage{graphicx}% Include figure files
\usepackage{dcolumn}% Align table columns on decimal point
\usepackage{bm}% bold math

\usepackage{comment}
\usepackage{amsthm}
\usepackage{amsmath}
\usepackage{amssymb}
\usepackage{graphicx}
\usepackage{url}
\usepackage{float}
\usepackage{bm}
\usepackage{ifthen}
\usepackage[usenames,dvipsnames]{color}
\usepackage{mathrsfs}
\usepackage[colorlinks=true,citecolor=blue,urlcolor=black]{hyperref}
\usepackage{float}
\usepackage{subfigure}
\usepackage{enumitem}
\usepackage{color}
\usepackage{setspace}
\usepackage{braket}

\usepackage{lipsum}
\newcommand{\be}{\begin{equation}}
	\newcommand{\ee}{\end{equation}}

\begin{document}
	
	\title{Prefixed-threshold Real-Time Selection for Free-Space Measurement-Device-Independent Quantum Key Distribution}
	
	%%% author
	\author{Wenyuan Wang}
	\affiliation{Centre for Quantum Information and Quantum Control (CQIQC), Dept. of Electrical \& Computer Engineering and Dept. of Physics, University of Toronto, Toronto,  Ontario, M5S 3G4, Canada}
	
	\author{Feihu Xu}
	%\affiliation{Research Laboratory of Electronics, Massachusetts Institute of Technology, 77 Massachusetts Avenue, Cambridge, Massachusetts 02139, USA}
	\affiliation{Shanghai Branch, National Laboratory for Physical Sciences at Microscale, University of Science and Technology of China, Shanghai, 201315, China}
	
	\author{Hoi-Kwong Lo}
	\affiliation{Centre for Quantum Information and Quantum Control (CQIQC), Dept. of Electrical \& Computer Engineering and Dept. of Physics, University of Toronto, Toronto,  Ontario, M5S 3G4, Canada}

	\begin{abstract}
		Measurement-Device-Independent (MDI) QKD eliminates detector side channels in QKD and allows an untrusted relay between two users. A desirable yet highly challenging application is to implement MDI-QKD through free-space channels. One of the major factors that affect the secure key rate in free-space MDI-QKD is atmospheric turbulence. In this work we show two important results: First, the independent fluctuations of transmittances in the two channels can significantly reduce MDI-QKD performance due to turbulence-induced channel asymmetry. Second, we consider the Prefixed Real-Time Selection (P-RTS) method we formerly introduced to BB84 and extend it to MDI-QKD. Users can monitor classical transmittances in their channels and improve performance by post-selecting signals in real-time based on pre-calculated thresholds. We show that we can establish a 2-dimensional threshold between Alice and Bob to post-select signals with both high signal-to-noise ratio and low channel asymmetry in real time, and greatly extend the maximum range of MDI-QKD in the presence of turbulence, which can be an important step towards future free-space MDI-QKD experiments.
	\end{abstract}

	\date{\today}
	\maketitle

\section{Background}

Quantum Key Distribution can theoretically provide unconditional security between two communicating users Alice and Bob. To address imperfections in practical laser sources, decoy-state protocol \cite{decoystate_LMC,decoystate_Hwang,decoystate_Wang} uses multiple levels of laser intensities to estimate secure single-photon components and thus avoiding photon-number-splitting attacks on multi-photons. 

However, practical devices - especially detector systems - are still susceptible to attacks. The Measurement-Device-Independent (MDI-QKD) QKD protocol \cite{mdiqkd} has been proposed to eliminate all detector side channels and hence preventing attacks on detectors. Here instead of Alice sending signals to Bob, they both send signals to a third-party Charles, who performs Bell-measurements on incoming signals and acts as an untrusted relay. Since its proposal, MDI-QKD has attracted much worldwide attention, and has seen many demonstrations in fibre systems \cite{mdiexp1,mdiexp2,mdiexperiment}, over as long as 404km of distance between users \cite{mdi404km}. MDI-QKD is also an ideal candidate in a quantum network since it allows untrusted relays, and there have been demonstrations of MDI-QKD network in a metropolitan setting \cite{mdinetwork}. Recently, MDI-QKD has also been extended to scenarios of asymmetric channels with different lengths \cite{mdi_asym} and demonstrated experimentally \cite{mdi_asym_exp}, which greatly extend its practicality by allowing the users to be placed at arbitrary locations.

\begin{figure}[h]
	\includegraphics[scale=0.47]{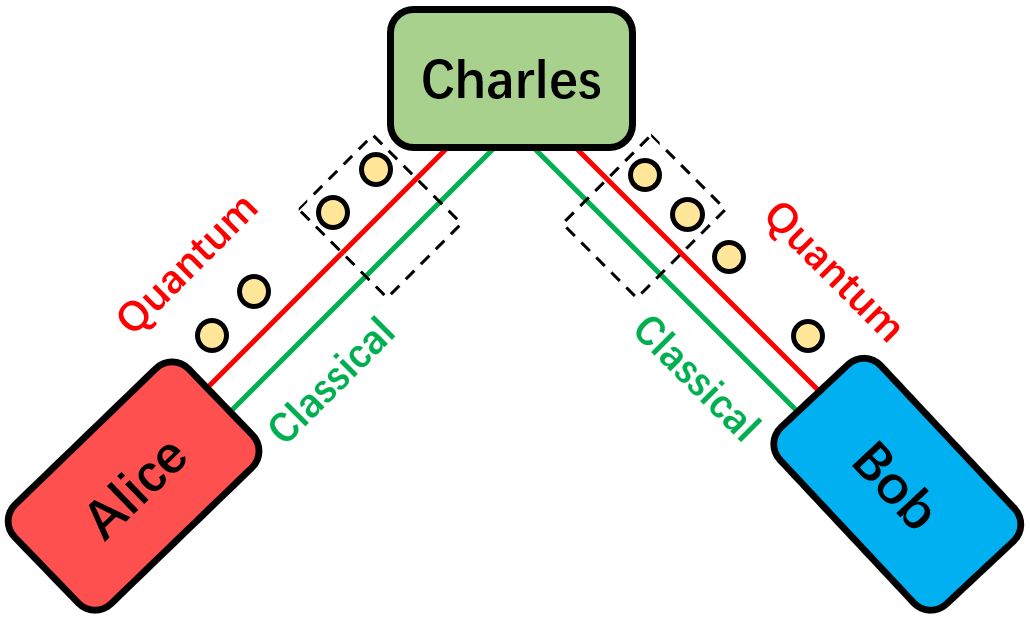}
	\caption{An illustration of MDI-QKD setup. Alice and Bob each sends quantum signals to a third-party Charles who acts as an untrusted relay. To perform real-time post-selection, Alice and Bob can each establish a classical channel alongside the quantum channel, to sample the channel transmittance in real time, and select only the sections where both channels have good transmittance. Note that, this classical channel could either be a strong laser at a slightly different wavelength, or it could be observables such as the count rate of the quantum detectors, which could also serve as indicator of channel transmittance.}
	\label{fig:setup}
\end{figure} 

A highly desirable application of MDI-QKD would be its implementation over free-space channels, which could allow mobile platforms such as ships, planes, satellites to communicate without detector susceptibilities It would also allow these users over moving platforms to join a dynamic quantum network with untrusted relays based on MDI-QKD. However, up to now, an experimental demonstration of free-space MDI-QKD remains challenging. In addition to the high level of loss in free-space channels, the atmospheric turbulence - which causes fluctuations in channel transmittances - also plays a large part in affecting the secure key rate. 

To address the high loss and strong turbulence in free-space channels, there have been proposals for post-selection methods on the signals, which can potentially increase the signal-to-noise ratio of the communication session. In \cite{probetest, SNRF}, it is proposed for BB84 protocol \cite{bb84} that one can sample the classical transmittance of a channel in real-time, and optimize a threshold for the recorded transmittance during post-processing, to discard low-transmittance signals and increase overall signal-to-noise ratio, hence obtaining higher key rate. In \cite{PRTS} we extended the idea in Ref. \cite{probetest} to decoy-state QKD, and also proposed that a fixed threshold can be \textit{pre-determined} before the experiment begins, independent of the channel condition, and users can use the threshold to select signals in real-time instead of waiting until the end of the session. This approach can reduce the storage requirements for Bob, and can also simplify post-processing computation since no optimization of threshold is needed.

In this work, we present two important results: Firstly, we show that in the presence of turbulence (scintillation of light, which causes transmittances to fluctuate in Alice's and Bob's channels), the key rate of MDI-QKD will decrease significantly due to turbulence-induced real-time asymmetry between the channels. This is contrary to many's popular belief and different from the results we observe for BB84 (where fluctuation of transmittance does not affect the key rate - and post-selection can make use of turbulence to increase key rate). We show that, without post-selection, key rate for MDI-QKD will drop significantly for turbulent channels. 

Secondly, we extend our P-RTS method to MDI-QKD, and show that by selecting a good threshold we can achieve much higher key rate and extended maximum tolerable channel loss. Moreover, our threshold does not depend on the channel condition and allows a semi-blind approach where ``bad" signals can be immediately discarded, which reduces storage and compute resource requirements. As atmospheric turbulence is very common in free-space channels, we believe that this work will be an important step towards the future experimental demonstration of MDI-QKD.

While this work is still under preparation, we notice a work on a similar subject published at \cite{QIP_MDI} and made online in December 2018. While it also applies post-selection to free-space MDI-QKD, importantly, it only uses a rather naive model of the problem that does not consider the turbulence-induced asymmetry (and assumes the key rate does not change in the presence of turbulence), which we show is a rather inaccurate overestimation of the key rate. Moreover, it suggests a simple ``square" threshold for post-selection (which needs optimization and depends on channel condition), while we show that Charles in fact has a much larger parameter space for threshold choice, and we propose a threshold that can closely approach optimality, and can be pre-determined prior to the experiment without the need to know the channel condition.

%summary of paper

\section{Theory}

In this section we define the models we use for the turbulent channel, for the post-selection, and also the models for a reliable estimation of the secure key rate. We point out an important point that, contrary to BB84 where fluctuation due to turbulence does not detrimentally affect key rate, in MDI-QKD even without post-selection, the key rate will decrease due to turbulence-induced channel asymmetry in real-time. To address this, we then propose solutions to set a good threshold for the post-selection. We will show in the next section with numerical results the effectiveness of the post-selection with our proposed thresholds.

\subsection{Channel Model under Turbulence}

%We have discussed applying real-time selection with a prefixed threshold in free-space BB84 QKD. 

In a free-space channel subject to atmospheric turbulence, the transmittance fluctuates with time and follows a probability distribution, which is often denoted as a probability distribution of transmission coefficient (PDTC) \cite{PDTC}. There are multiple models for such a PDTC function, a commonly used model is the log-normal distribution \cite{laser}:

\begin{equation}
p_{\eta_0,\sigma}(\eta)={1 \over {\sqrt{2\pi}} \sigma\eta}e^{-{{[ln({\eta \over \eta_0})+{1\over 2}{\sigma}^2]^2}\over{2\sigma^2}}}
\end{equation}

\noindent where the channel is described by two parameters $(\eta_{0},\sigma)$ that respectively represent the mean transmittance and the variance of the channel. The log-normal distribution satisfies normalization condition when $\eta$ is small

\begin{equation}
	\int_{0}^{1}p_{\eta_0,\sigma}(\eta)d\eta \approx 1
\end{equation}

\noindent When $\eta$ is comparable to 1, there is a non-negligible portion of the probability that $\eta>1$, in this case we should calculate the integral from 0 to 1 to obtain a constant normalization factor (smaller than 1), and divide the PDTC with this factor, i.e. forming a truncated log-normal distribution.

By post-selecting $\eta$ with a threshold $\eta_T$, we can have a higher average transmittance among post-selected signals:

\begin{equation}
\langle \eta \rangle={{\int_{\eta_T}^{1}\eta p_{\eta_0,\sigma}(\eta)d\eta}\over{\int_{\eta_T}^{1}p_{\eta_0,\sigma}(\eta)d\eta}}
\end{equation}

\noindent Again, the post-selected signals follow a truncated log-normal distribution between $[\eta_T,1]$, hence a normalization factor (total probability within the post-selected region) is included.\\

\begin{figure*}[t]
	\includegraphics[scale=0.19]{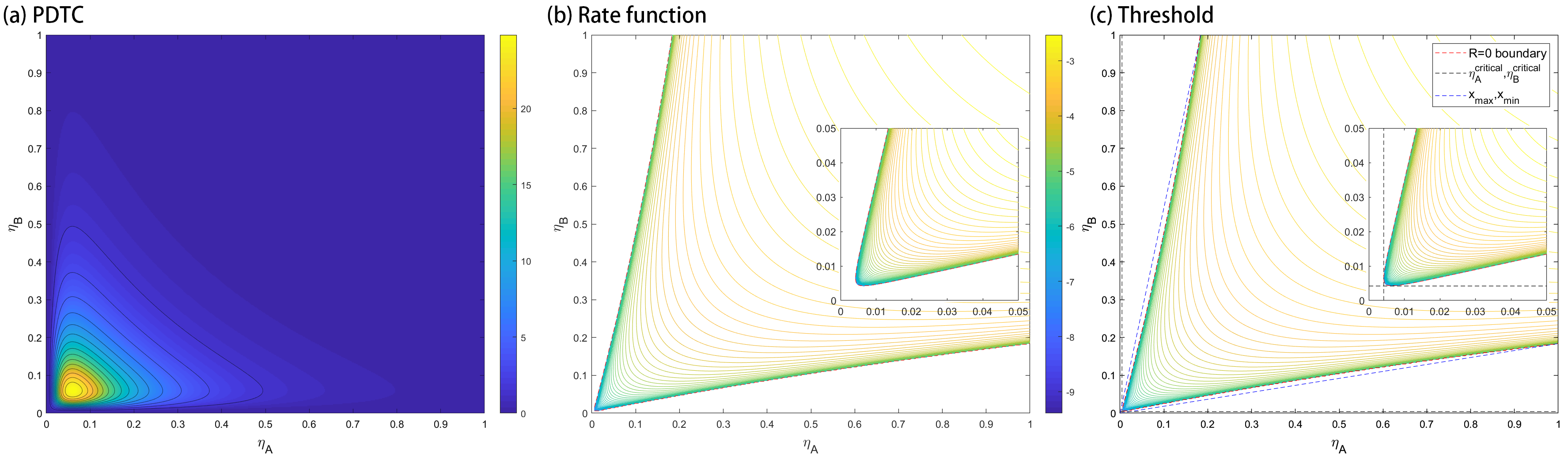}
	\caption{(a) A joint PDTC function for Alice's and Bob's real-time transmittances. (b) The contours of key rate function $R(\eta_A,\eta_B)$ for 4-intensity MDI-QKD protocol, for fixed intensities and infinite data size (plotted in $log_{10}$ scale). We can see that the $R=0$ contour follows a near-hyperbolic shape, whose asymptotic lines represent the maximum and minimum acceptable channel mismatch $x={\eta_A/\eta_B}$. There is also a ``gap" near the origin mainly determined by the noise (e.g. dark counts). (c) Choice of thresholds. We can choose the R=0 boundary as the threshold. Or for simplicity, we can also approximate it using the horizontal/vertical tangents $\eta_A^{critical},\eta_A^{critical}$ of the $R=0$ contour (which discard signals with low signal-to-noise ratio), combined with the asymptotic lines $x_{max},x_{min}$ (which discard signals with strong channel asymmetry). Importantly, all the information in this plot comes from the structure of $R(\eta_A,\eta_B)$ and are independent of the actual channel PDTC, i.e. they can be pre-determined before the experiment. In this plot $\eta_{A}^{critical}=\eta_{B}^{critical}=0.0042$, $x_{min}=1/x_{max}=0.184$.}
	\label{fig:PDTC}
\end{figure*}

Now, let us consider MDI-QKD, where Alice and Bob are each connected to Charles with a channel. Intuitively, here we can first assume that both channels are subject to atmospheric turbulence, and that their fluctuations are \textit{independent and non-correlated}. We can denote the channel transmittances as $\eta_A,\eta_B$ respectively. Then, the joint PDTC for the two channels can be written as

\begin{equation}
p_{AB}(\eta_A,\eta_B)=p_{\eta_{A_0},\sigma_A}(\eta_A)\times p_{\eta_{B_0},\sigma_B}(\eta_B)
\end{equation}

\noindent where the two channels are described by the $(\eta_{A_0},\sigma_A)$, $(\eta_{B_0},\sigma_B)$ channel condition parameters. The joint PDTC also follows the normalization condition:
\begin{equation}
\begin{aligned}
&\iint p_{AB}(\eta_A,\eta_B)d\eta_A d\eta_B\\
&= \int_{0}^{1}p_{\eta_{A_0},\sigma_A}(\eta_A)d\eta_A \times \int_{0}^{1}p_{\eta_{B_0},\sigma_B}(\eta_B)d\eta_B\\
&\approx 1
\end{aligned}
\end{equation}

The joint PDTC can be considered as a two-variable function on a plane defined by $(\eta_A,\eta_B)$, as shown in Fig.\ref{fig:PDTC} (a). Now, we observe that for a post-selection on the signals received by Charles, he can actually observe both Alice's and Bob's channel transmittances $(\eta_A,\eta_B)$, and make a decision based on these two observables. Importantly, he does not have to independently set a threshold for each channel respectively (and select events where both transmittances pass the threshold), but rather, he is able to make a joint decision based on the two observables - for instance, selecting events based on a high level of symmetry between $(\eta_A,\eta_B)$ is present, instead of based on the respective signal strength of $\eta_A,\eta_B$ alone. Mathematically, Charles is selecting a domain $\Omega \subseteq \bm{R}^2$ in the 2D space defined by $(\eta_A,\eta_B)$.

In the selected domain, we can perform a 2D integral to obtain the expected values of the transmittances.

\begin{equation}
\begin{aligned}
\langle \eta_A \rangle&={{ \iint_{\Omega}\eta_A p_{AB}(\eta_A,\eta_B)d\eta_A d\eta_B}\over{ \iint_{\Omega}p_{AB}(\eta_A,\eta_B)d\eta_A d\eta_B }}\\
\langle \eta_B \rangle&={{ \iint_{\Omega}\eta_B p_{AB}(\eta_A,\eta_B)d\eta_A d\eta_B}\over{ \iint_{\Omega}p_{AB}(\eta_A,\eta_B)d\eta_A d\eta_B }} 
\end{aligned}
\end{equation}

In the simple case of a "square" threshold, i.e. 

\begin{equation}
	\Omega^{square} = \{(\eta_A,\eta_B)\in \bm{R}^2: \eta_{A_T}\leq \eta_A \leq 1, \eta_{B_T}\leq \eta_B \leq 1\}
\end{equation}

\noindent the post-selection follows two independent thresholds $\eta_{A_T},\eta_{B_T}$. This is the simplest form of threshold Charles can implement, and the probability distribution can be decoupled between $\eta_A$ and $\eta_B$, hence one can simply use Eq. (3) to calculate the mean transmittances. However, note that there are more careful (and potentially better) ways to select such a threshold, to make use of Charles' joint knowledge of $(\eta_A,\eta_B)$, which we will discuss in later sections.

\subsection{Models for Key Rate}

In \cite{PRTS}, for BB84 protocol with a single free-space channel, we have proposed two models:
\begin{equation}
\begin{aligned}
R^{\text{Simplified}}_{BB84}(\eta_T) &= \left(\int_{\eta_T}^{1}p_{\eta_0,\sigma}(\eta)d\eta\right) \times R(\langle \eta \rangle) \\
R^{\text{Integration}}_{BB84} &=\int_{0}^{1}R( \eta )p_{\eta_0,\sigma}(\eta)d\eta
\end{aligned}
\end{equation}

\noindent where $R(\eta)$ is the key rate function (where all other experimental parameters are fixed, e.g. dark count, detector efficiency, misalignment etc.), and $\eta_T$ is the threshold used to post-select signals according to the real-time transmittance. The ``\textbf{simplified model}" $R^{\text{Simplified}}_{BB84}(\eta_T)$ finds the mean transmittance among the post-selected signals, and calculates the key rate with a static model with this new transmittance, i.e. it assumes that all the signals are transmitted with this mean transmittance. It is also multiplied by the proportion of selected signals (since the total number of signals decreases due to post-selection). On the other hand, the ``\textbf{rate-wise integration model}" $R^{\text{Integration}}_{BB84}$ (for simplicity in the following text we will just call it integration model in short) divides all signals into bins of $[\eta,\eta+\Delta \eta)$ and adds up the key rate in all bins. In the asymptotic (infinite-data) limit, the integration model can make use of the entire probability distribution's information, and always produces higher key rate than the simplified model. Effectively, it provides an upper-bound to the maximum key rate $R^{\text{Simplified}}_{BB84}(\eta_T)$ can achieve by adjusting the threshold $\eta_T$:

\begin{equation}
R^{\text{Simplified}}_{BB84}(\eta_T) \leq R^{\text{Integration}}_{BB84}
\end{equation}

For BB84, the near-linearity of the rate function $R(\eta)$ guarantees a fixed optimal threshold $\eta_{critical}$ exists, where $R=0$ for all $\eta \leq \eta_{critical}$. This optimal threshold position is calculated with $R(\eta)$ only, and is independent of the channel condition $(\eta_0, \sigma)$. Hence, we can find a prefixed threshold $\eta_{critical}$ that maximizes the performance of BB84 with post-selection, satisfying:

\begin{equation}
	R^{\text{Simplified}}_{BB84}(\eta_{critical}) = R^{\text{Integration}}_{BB84}
\end{equation}\\

Now, for MDI-QKD, we can firstly extend the concepts in the BB84 case, and define the simplified model and the integration model as following:
\begin{equation}
\begin{aligned}
R^{\text{Simplified}}(\Omega) &= R(\langle \eta_A \rangle,\langle \eta_B \rangle) \\
&\times \left( \iint_{\Omega}p_{AB}(\eta_A,\eta_B)d\eta_A d\eta_B \right) \\
R^{\text{Integration}} &= \int_0^1 \int_0^1 R(\eta_A,\eta_B) p_{AB}(\eta_A,\eta_B)d\eta_A d\eta_B
\end{aligned}
\end{equation}

However, a crucial point here is that, for MDI-QKD, the simplified model \textit{does not} accurately represent the key rate. The reason is that MDI-QKD heavily depends on the \textit{symmetry} between channel transmittances (because it makes use of a two-photon interference in the X basis, and its quantum bit error rate (QBER) will depend on the interference visibility). Suppose the mean transmittances in the channels $\eta_{A_0},\eta_{B_0}$ are equal, and Alice and Bob choose the same intensities. Then, without post-selection, we can obtain $\langle \eta_A \rangle = \eta_{A_0}, \langle \eta_B \rangle=\eta_{B_0}$. This means that, when calculating the simplified model based on $R(\langle \eta_A \rangle,\langle \eta_B \rangle)$, we are assuming a perfectly symmetric setup (which will presumably result in low QBER in the X basis and a high estimated key rate). However, in reality, $\eta_A,\eta_B$ are independent variables, and they are very likely not equal in real-time for the majority of times. This means that, any deviation from $\eta_A=\eta_B$ will result in an increase in the QBER in the X basis. Overall, when one collects the observables (counts and error-counts), he/she will find a much larger-than-expected QBER in the X basis, preventing him/her from acquiring good estimation of the phase-error rate and a good key rate. Therefore, simplified model \textit{overestimates} the key rate.

In other words, we make the observation that turbulence-induced channel asymmetry in real-time will decrease the key rate of MDI-QKD. This is very different from what we observed for BB84, where key rate, gain, and error-gain are all near-linear functions, and any increase/decrease in error-gain due to fluctuations cancel out when computing the mean value. On the other hand, for a pair of channels with symmetric mean transmittances, fluctuation always decreases the visibility and increases the QBER.

Since simplified model is not an accurate model anymore for MDI-QKD, here we propose a better representation of the key rate in turbulence. Consider the process of obtaining key rate for MDI-QKD, for instance, for the 4-intensity MDI-QKD protocol \cite{mdifourintensity}:
\begin{equation}
\begin{aligned}
R=P_{s}^2 \{(s e^{-s})^2 Y_{11}^{X,L}[1-h_2(e_{11}^{X,U})]\\
-f_eQ_{ss}^Z h_2(E_{ss}^Z)\}
\end{aligned}
\end{equation}

%(as an example let us consider the 4-intensity protocol \cite{mdifourintensity} where X and Z bases are decoupled). The users will first collect \texit{observables} in the X and Z bases, namely $Q_ij$

\begin{figure*}[t]
	\includegraphics[scale=0.22]{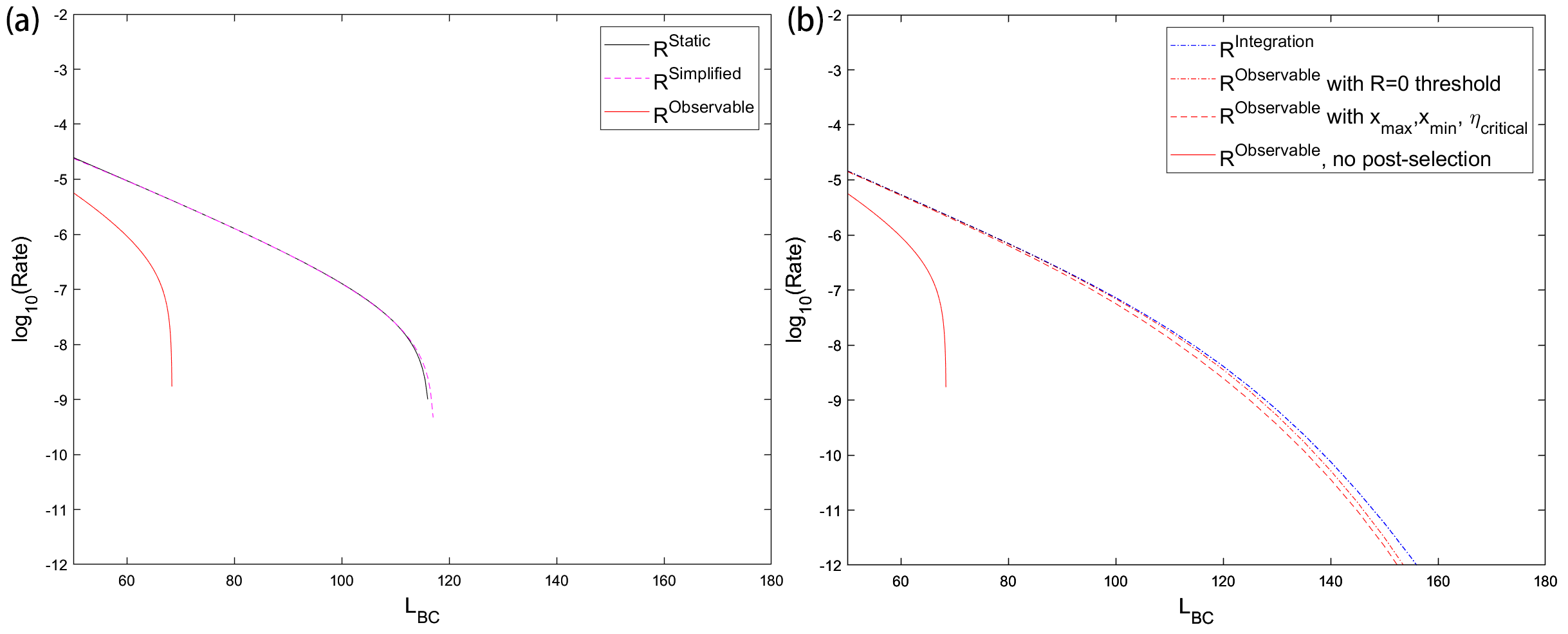}
	\caption{(a) Comparison of key rate models without post-selection. As can be seen, the simplified model incorrectly assumes the same rate as a static model, while from the observable model we can see that MDI-QKD key rate decreases significantly with turbulence. (b) Comparison of key rate obtained with different thresholds. As can be seen, both the R=0 boundary and the straight-line approximations ($x_{min},x_{max},\eta_A^{critical},\eta_B^{critical}$) greatly increase key rate and maximum distance/loss in the channel. Here for convenience we've plotted the key rate versus distance between Charles and Bob in standard optical fibre (0.2dB/km), but in terms of dB we can also see over 30dB of maximum increase in channel loss between Alice and Bob.}
	\label{fig:rate}
\end{figure*} 
\noindent This protocol uses one signal intensity $s$ for the Z basis to generate the key, and three decoy intensities each for Alice and Bob $\{\mu,\nu,\omega\}$ to perform decoy-state analysis in the X basis. Here the constants include the intensities and the probabilities of sending them, such as $s,P_{s}$, and the error-correction efficiency $f_e$. We can see that the ``variables" that change with $\eta_A,\eta_B$ are the observed gain and QBER in the Z basis $Q_{ss}^Z, E_{ss}^Z$, and the single-photon contributions $Y_{11}^{X,L},e_{11}^{X,U}$ estimated from the X basis observables $Q_{ij}^X, E_{ij}^X$ where $i,j \in \{ \mu,\nu,\omega \}$. Overall, the key rate can be considered as a function of the observables:

\begin{equation}
\begin{aligned}
R(\eta_A,\eta_B) = R[&Q_{ij}^X(\eta_A,\eta_B), T_{ij}^X(\eta_A,\eta_B),\\
&Q_{ss}^Z(\eta_A,\eta_B), T_{ss}^Z(\eta_A,\eta_B)]
\end{aligned}
\end{equation}

\noindent here $T_{ij}^X = Q_{ij}^X E_{ij}^X$ are the error-gains (which correspond to the actual observed error-counts) in the X basis. Similar goes for $T_{ss}^Z=Q_{ss}^Z E_{ss}^Z$. All the error-gains and gains are functions of $\eta_A,\eta_B$ too.

In an actual experiment, the users collect the corresponding counts and error-counts over the entire session, and divide them by the number of signals sent respectively for each intensity combination to acquire the average gain and error-gain. The X basis gain and error-gain are used in decoy-state analysis for privacy amplification, and the Z basis error-counts and counts follow error-correction and key generation. We can see that, since an experiment actually performs privacy amplification and error-correction on the \textit{observables} collected over the entire session and calculates their average values, we can define an \textbf{observable model} that accurately represents the expected key rate in experiment:

\begin{equation}
\begin{aligned}
R^{\text{Observable}}(\Omega) = R[&\langle Q_{ij}^X \rangle, \langle T_{ij}^X \rangle, \langle Q_{ss}^Z \rangle, \langle T_{ss}^Z \rangle]
\end{aligned}
\end{equation}

\noindent This model represents the actual observables one would get in an experiment, and takes into consideration the effect turbulence-induced fluctuations can have on the average QBER (error-counts) in the X basis. We plot the observable model versus static and simplified model in Fig. \ref{fig:rate} (a) (without applying any post-selection). We can see that simplified model fails to characterize the effect of turbulence and assumes the same key rate as a static channel, while observable model shows that MDI-QKD key rate greatly decreases with turbulence if not addressed actively.

\subsection{Choice of Post-selection Thresholds}

In this subsection we discuss some good choices for the threshold $\Omega$ Charles uses when post-selecting signals.

Let us first plot out the key rate versus $\eta_A,\eta_B$ function in Fig.\ref{fig:PDTC} (b). Note that this function is only determined by the experimental parameters (misalignment, dark count, detector efficiency) and the intensities Alice and Bob choose, and it \textit{doesn't depend on} the joint PDTC of the channels. 

From it we can see that the key rate follows a near-parabolic shape, with two asymptotic lines corresponding to the maximum and minimum channel asymmetry $x=\eta_A/\eta_B$ (which graphically correspond to the reciprocal of slope). This is reasonable because the QBERs (mainly $E_{ij}^X$, and $E_{ss}^Z$ which is less sensitive but still affected) depend on the channel asymmetry, and the key rate becomes zero at two cut-off points $x_{max},x_{min}$. The existence of these two cutoff lines for asymmetry are analytically proven for the infinite-decoy case in Ref.\cite{mdi_asym}, which shows that for $R=0$ there are two groups of solutions at $x_{max},x_{min}$, regardless of the actual amplitudes of $\eta_A,\eta_B$, while for the case of finite-decoys the result is numerically shown (although yet to be proven analytically because there is no simple analytical formula for $E_{ij}^X$).

In BB84, the optimal threshold we select was $\eta_{critical}$ such that all $\eta<\eta_{critical}$ satisfy $R(\eta)=0$. Similarly, for MDI-QKD, since we know all the information in the $R(\eta_A,\eta_B)$ plot, here we can propose to use a threshold $\Omega^{boundary}$ defined by where $ R(\eta_A,\eta_B) \geq 0$:

\begin{equation}
\Omega^{boundary} = \{(\eta_A,\eta_B)\in \bm{R}^2: R(\eta_A,\eta_B) \geq 0 \}
\end{equation}

To simplify the implementation, it's also possible to approximate this boundary (which takes a near-parabolic shape) with four straight lines, representing two characteristics:  $\eta_{A}^{critical},\eta_{B}^{critical}$ (which are mainly determined by the dark counts), and $x_{max},x_{min}$ (which are mainly affected by basis misalignment). We can then require the signals to jointly satisfy the conditions on signal-to-noise ratio and symmetry, i.e.

\begin{equation}
\begin{aligned}
\Omega^{joint} = \{(\eta_A,\eta_B)\in \bm{R}^2: &\eta_{A_T}\leq \eta_A \leq 1, \eta_{B_T}\leq \eta_B \leq 1, \\
&x_{min} \leq \eta_A/\eta_B \leq x_{max} \}
\end{aligned}
\end{equation}

\noindent The two thresholds are plotted in Fig. \ref{fig:PDTC}(c). Importantly, the plot $R(\eta_A,\eta_B)$ is generated without any information of the PDTC, and all the above information including $R=0$ boundary, $\eta_{A}^{critical},\eta_{B}^{critical}$, and $x_{max},x_{min}$ are all acquired from the plot of $R(\eta_A,\eta_B)$ alone, which only depends on the intensities and the experimental parameters (misalignment, dark count rate, detector efficiency etc.), but are independent of the actual joint PDTC of the channel. This means that they can be very conveniently pre-determined before the experiment for real-time post-selection of signals (instead of having to optimize the threshold after the experiment).\\

%\begin{equation}
%\Omega^{boundary} = \{(\eta_A,\eta_B)\in \bm{R}^2: R(\eta_A,\eta_B) \geq 0 \}
%\end{equation}

\section{Numerical Results}

Here we present a numerical simulation for the post-selection method we proposed. For simplicity, here we consider a 4-intensity protocol with infinite-data size and fixed intensities.
First, we tested the key rate of MDI-QKD in the presence of turbulence without post-selection. As can be seen, the simplified model ``overestimates" the key rate since it assumes the same key rate as a static model. The observable model correctly captures the decrease in key rate due to turbulence-induced asymmetry. We can see that, without post-selection, the performance of MDI-QKD greatly decreases in the presence of turbulence.

In Fig. \ref{fig:rate}(b), we plot the observable model obtained from the two threshold methods versus no post-selection. As can be seen, the $R=0$ boundary $\Omega^{boundary}$ (and the approximated $\Omega^{joint}$ which has a key rate just slightly less than the former) captures the most information of the key rate function and results in a key rate very similar to the upper bound of integration model $R^{\text{Integration}}$, which asymptotically utilizes all information of the probability function and is expected to always produces higher key rate than models that utilize average values - i.e. an upper-bound for the observable model. Nonetheless, we can see that using the thresholds, the performance we obtain with observable model approaches this upper bound very closely, meaning that the thresholds we propose are near-optimal.\\

\section{Discussions}

In this work we make an important observation that turbulence-induced channel asymmetry decreases the key rate of MDI-QKD. This means that using post-selection is not only a potential means of improvement, but actually might be necessity in acquiring a good rate, since simply not addressing the turbulence will result in low rate. We then proposed the powerful solution of a prefixed-threshold post-selection method for MDI-QKD, that can greatly increase the maximum tolerable loss in MDI-QKD communications, paving the way towards future experimental implementations of MDI-QKD.

A few remaining questions include: (1) although we numerically show that certain threshold choices can result in near-optimal key rate, it remains to be shown analytically why such a boundary of $R=0$ gives maximum key rate after post-selection. (2) In reality the free-space channels between Alice and Bob are likely not of equal mean transmittances (e.g. due to different distances). The $R(\eta_A,\eta_B)$ map (and R=0 boundary) method in principle holds true for asymmetric cases where the two channels have different mean transmittances $\eta_{A_0} \neq \eta_{B_0}$ (which will be represented by a lopsided joint PDTC), and the users can also choose different intensities (which results in a lopsided $R(\eta_A,\eta_B)$ contour too). More testing remains to be shown for these cases. (3) finite-size effects are important considerations in practical QKD. For this case the intensities (and the probabilities of sending them) need to be highly optimized for a good key rate, and the optimal parameters change with distance - which changes the rate function too. Moreover, post-selection not only decreases the total amount of signals, but will also result in stronger statistical fluctuation among post-selected signals too, hence affecting the key rate. In this case a more careful discussion is needed. These questions will be the subject of our future studies.


\begin{thebibliography}{}
	
	\bibitem{decoystate_Hwang} WY Hwang, Phys. Rev. Lett. 91.5 (2003): 057901.
	
	\bibitem{decoystate_LMC} HK Lo, X Ma, K Chen, Phys. Rev. Lett. 94.23 (2005): 230504.
	
	\bibitem{decoystate_Wang} XB Wang, Phys. Rev. Lett. 94.23 (2005): 230503.
	
	\bibitem{mdiqkd} HK Lo, M Curty, B Qi, Phys. Rev. Lett. 108.13 (2012): 130503.
	
	\bibitem{mdiexp1} TF da Silva, D Vitoreti, GB Xavier, GC do Amaral, GP Tempor\~ao, JP von der Weid, Phys. Rev. A 88, 052303 (2013)
	
	\bibitem{mdiexp2}  Y Liu et al., Phys. Rev. Lett., 111:130502, (2013).
	
	\bibitem{mdiexperiment} Z Tang, Z Liao, F Xu, B Qi, L Qian, HK Lo, Phys. Rev. Lett. 112.19, 190503 (2014).
	
	
	\bibitem{mdi404km} HL Yin et al., Phys. Rev. Lett. 117.19, 190501 (2016).
	\bibitem{mdinetwork} YL Tang et al., Phys. Rev. X 6.1, 011024 (2016).
	
	\bibitem{mdi_asym} W Wang, F Xu, HK Lo, arXiv preprint arXiv:1807.03466 (2018).
	
	\bibitem{mdi_asym_exp} H Liu, et al., arXiv preprint arXiv:1808.08584 (2018).
	
	\bibitem{probetest} G Vallone et al. Phys. Rev. A 91, 042320 (2015).
	
	\bibitem{SNRF} C Erven et al. New Journal of Physics 14.12 (2012): 123018.
	\bibitem{bb84} C Bennett, G Brassard, International Conference on Computer System and Signal Processing, IEEE (1984).
	\bibitem{PRTS} W Wang, F Xu, HK Lo, Phys. Rev. A 97.3 (2018): 032337.
	\bibitem{QIP_MDI} ZD Zhu, et al. Quant. Info. Proc. 18.1 (2019): 33.
	\bibitem{PDTC} A Semenov, W Vogel, Phys. Rev. A 80, 021802(R) (2009).
	
	\bibitem{laser} LC Andrews, RL Phillips, CY Hopen, Vol. 99. SPIE press (2001).
	

	
	%\bibitem{mdipractical} Xu, Feihu, et al. "Practical aspects of measurement-device-independent quantum key distribution." New Journal of Physics 15.11 (2013): 113007.s
	
	%\bibitem{mdiPOP} Rubenok, Allison, et al. "Real-world two-photon interference and proof-of-principle quantum key distribution immune to detector attacks." Physical review letters 111.13 (2013): 130501.
	
	%\bibitem{probetest} G Vallone et al {\em Adaptive real time selection for quantum key distribution in lossy and turbulent free-space channels} Phys. Rev. A 91, 042320 (2015)
	
	%\bibitem{mdiparameter} Xu, Feihu, He Xu, and Hoi-Kwong Lo. "Protocol choice and parameter optimization in decoy-state measurement-device-independent quantum key distribution." Physical Review A 89.5 (2014): 052333.
	
	\bibitem{mdifourintensity} YH Zhou, ZW Yu, XB Wang, Phys. Rev. A 93.4 (2016): 042324.
	


	
		

	

\end{thebibliography}
\end{document}